\documentclass[journal=jctcce,manuscript=communication]{achemso}
\setkeys{acs}{maxauthors=0,articletitle=true}

\usepackage{achemso}
\usepackage{graphics}
\usepackage{amssymb,amsfonts}
\usepackage{graphicx}
\usepackage[table]{xcolor}
\usepackage{multirow}
\usepackage{caption}
\usepackage{subcaption}
\usepackage{booktabs}
\usepackage{colortbl}
\usepackage{amsmath}
\usepackage{amsopn}
\usepackage{bm}
\usepackage{siunitx}
\usepackage{bm}
\usepackage{color}
\usepackage{array}
\usepackage{lscape}
\usepackage{mciteplus}
\usepackage[version=3]{mhchem}
\usepackage{ulem}
\usepackage{listings}
\usepackage{enumerate}
\usepackage{lmodern}

\SectionNumbersOn

\author{Janus J. Eriksen}
\email{janus.eriksen@bristol.ac.uk}
\affiliation[University of Bristol]
{School of Chemistry, University of Bristol, Cantock's Close, Bristol BS8 1TS, United Kingdom}

\title[TITLE]{The Shape of Full Configuration Interaction to Come}

\begin{document}

%
%%%%%%%%%
%  ABSTRACT  %
%%%%%%%%%
%
\begin{abstract}

We present a Perspective on what the future holds for full configuration interaction (FCI) theory, with an emphasis on conceptual rather than technical details. Upon revisiting the early history of FCI, a number of its key contemporary approximations are compared on as equal a footing as possible, using a recent blind challenge on the benzene molecule as a testbed [Eriksen {\textit{et al.}}, {\textit{J. Phys. Chem. Lett.}}, $\bm{11}$, 8922 (2020)]. In the process, we review the scope of applications for which FCI continues to prove indispensable, and the required traits in terms of robustness, efficacy, and reliability its modern approximations must satisfy are discussed. We close by conveying a number of general observations on the merits offered by the state-of-the-art alongside some of the challenges still faced to this day. While the field has altogether seen immense progress over the years---the past decade, in particular---it remains clear that our community as a whole has a substantial way to go in enhancing the overall applicability of near-exact electronic structure theory for systems of general composition and increasing size.

\end{abstract}

\newpage

%
%%%%%%%%%%%%
%    TOC GRAPHIC
%%%%%%%%%%%%

%
\section*{TOC Graphic}
\begin{figure}[ht]
\begin{center}
\includegraphics[width=\textwidth]{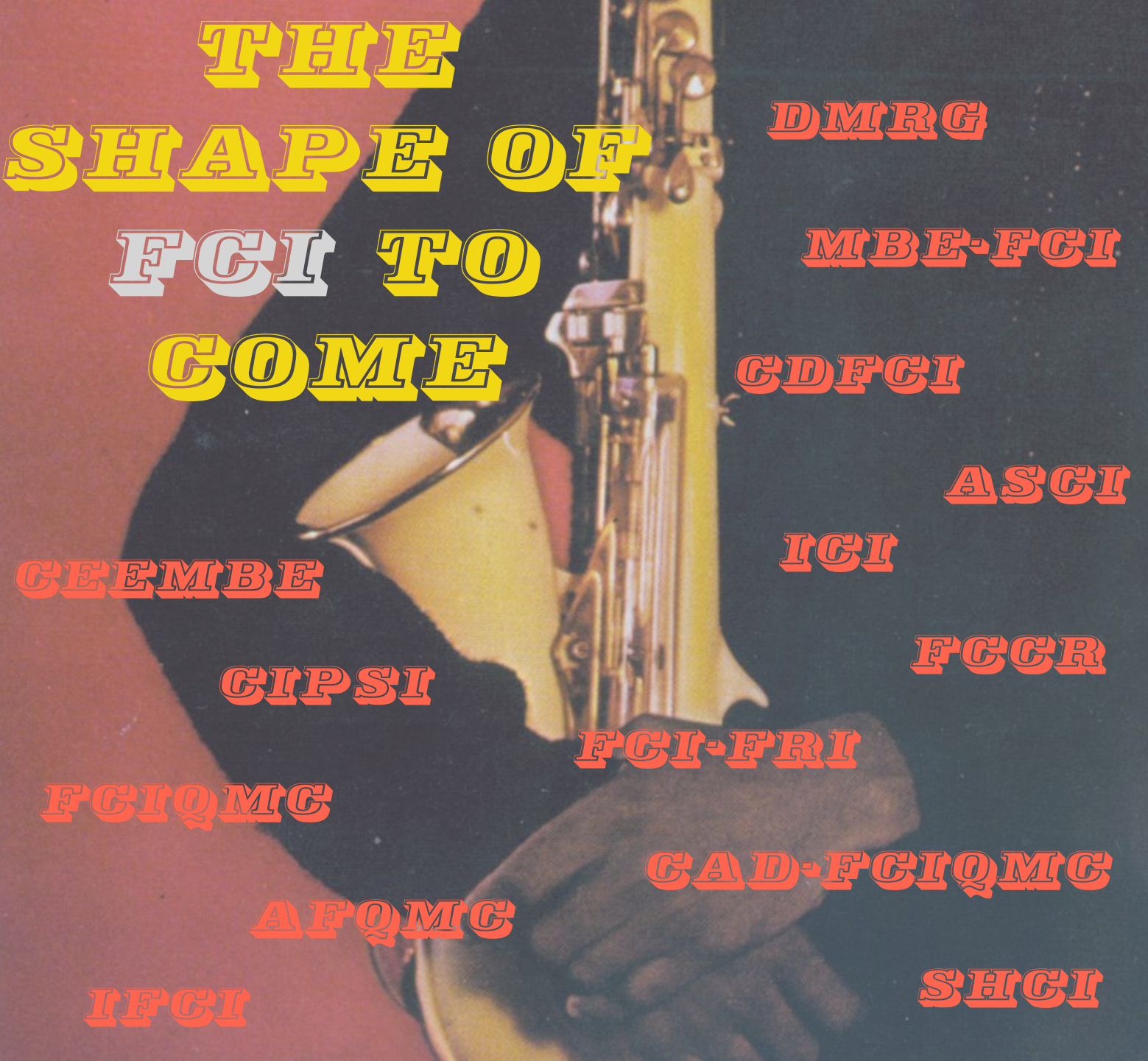}
\label{toc_fig}
\end{center}
\end{figure}

\newpage

%
%%%%%%%%%%%
%  INTRODUCTION
%%%%%%%%%%%
%

Drawing on earlier work on direct configuration interaction algorithms~\cite{handy_fci_cpl_1980,siegbahn_fci_cpl_1984,knowles_handy_fci_cpl_1984}, Knowles and Handy first reported---in 1989---on a truly impressive new implementation of full configuration interaction (FCI) theory capable of performing calculations of unlimited size ({\textit{sic}})~\cite{knowles_handy_fci_jcp_1989}. The following year, Olsen and co-workers succeeded in passing the billion determinant limit~\cite{olsen_fci_cpl_1990}, and the combined achievements of these two groups~\cite{knowles_handy_fci_cpc_1989,knowles_fci_cpl_1989,olsen_fci_jcp_1988} were long regarded by many in the community as heralding a new age of exact electronic structure theory, spawning a number of additional developments in the area over the following years~\cite{zarrabian_sarma_paldus_fci_cpl_1989,harrison_zarrabian_fci_cpl_1989,evangelisti_fci_jcp_1993,evangelisti_fci_jcc_1998,sherrill_fci_adv_quant_chem_1999,gordon_kendall_fci_jcp_2003,gan_harrison_fci_ieee_2005}. Since the 1990s, however, much of the optimism surrounding standard FCI has come to fade. While boundaries continue to be pushed today~\cite{fales_levine_gpu_fci_jcp_2015,vogiatzis_parallel_mcscf_fci_jcp_2017}, it has now become abundantly clear that truly exact theory will never successfully evolve into a widely applicable commodity tool, at least not by means of classical computing~\cite{aspuru_guzik_head_gordon_quant_comput_science_2005,aspuru_guzik_white_quant_comput_nat_chem_2010,peruzzo_aspuru_guzik_o_brien_quant_comput_nat_commun_2014,kivlichan_babbush_quant_comput_prl_2018,babbush_chan_quant_comput_prx_2018,aspuru_guzik_quant_comput_chem_rev_2019,chan_quant_comput_chem_rev_2020}, and the emergence of its quantum counterpart is not guaranteed to offer a practical panacea either~\bibnote{On a quantum computer, measurements of the energy and projections to an {\textit{arbitrary}} eigenstate, by phase estimation, is efficient. However, the preparation of a state with sufficient overlap with a {\textit{particular}} state, e.g., the exact ground state, in a way such that phase estimation yields a specific eigenstate, will generally be exponential in cost. For further details on this, see, for instance, the recent review in Ref. \citenum{chan_quant_comput_chem_rev_2020}}. In its benchmarking capacity, on the other hand, modern FCI has inevitably been transformative in helping to manifest theoretical chemistry as a scientific discipline with unquestionable interpretative and predictive powers~\cite{graham_bartlett_fci_jcp_1986,bauschlicher_taylor_fci_jcp_1986,bauschlicher_knowles_fci_jcp_1986,bauschlicher_taylor_dipole_fci_tca_1987,bauschlicher_langhoff_trans_dipole_fci_tca_1988,bauschlicher_langhoff_fci_lif_jcp_1988,bauschlicher_langhoff_fci_alh_jcp_1988,bauschlicher_fci_tih_jpc_1988,koch_harrison_fci_jcp_1991,knowles_werner_fci_cpl_1991,evangelisti_fci_cpl_1999,evangelisti_gagliardi_fci_cp_1994,evangelisti_fci_jcp_2004,evangelisti_fci_jcp_2009,evangelisti_paulus_fci_jcp_2008,van_lenthe_fci_jcp_1995,christiansen_fci_cpl_1996,olsen_bond_break_h2o_jcp_1996,halkier_gauss_dipole_fci_jcp_1999,larsen_gauss_polarizability_fci_jcp_1999,olsen_bond_break_n2_jcp_2000,olsen_sundholm_fci_quant_dot_prb_2000,olsen_sundholm_fci_quant_dot_mp_2002,larsen_olsen_jorgensen_fci_jcp_2001,thogersen_olsen_fci_cpl_2004,sherrill_ch2_jcp_1998,sherrill_bond_break_hf_jcp_2003,abrams_sherrill_fci_jcp_2003,abrams_sherrill_bond_break_c2_jcp_2004,sherrill_piecuch_bond_break_c2_jcp_2005,sherrill_ct_fci_jcp_2007,auer_gauss_pecul_fci_cpl_2003,harrison_dixon_fci_jcp_2006,marin_velasco_fci_jcc_2008,martin_fci_jcp_2008,aquilanti_fci_cpl_2009,schaefer_fci_mp_2013}. In addition, the availability of versatile FCI codes has been a prime driving force in the realization of high-order many-body perturbation (MBPT) and coupled cluster (CC) theories~\cite{knowles_handy_mp_cpl_1985,handy_knowles_mp_tca_1985,handy_knowles_mp_jpc_1988,olsen_general_order_cc_jcp_2000,kallay_string_based_cc_jcp_2001}, studies of their possible divergences~\cite{nobes_handy_knowles_mp_cpl_1987,mp_divergence_olsen_jcp_1996,mp_divergence_christiansen_cpl_1996,mp_divergence_halkier_jcp_1999,mp_divergence_olsen_jcp_2000,mp_divergence_larsen_jcp_2000,mp_divergence_sherrill_jcp_2000,eriksen_e_ccsd_tn_jcp_2016,eriksen_convergence_ccpt_jcp_2016,pert_theory_olsen_jcp_2019}, and as the principal component of complete active space self-consistent field~\cite{roos_casscf_acp_1987,roos_casscf_benzene_cpl_1992,olsen_casscf_ijqc_2011,zgid_nooijen_dmrg_rdms_jcp_2008_1,zgid_nooijen_dmrg_rdms_jcp_2008_2,ghosh_chan_dmrg_casscf_jcp_2008,yanai_morokuma_dmrg_casscf_jctc_2013,nakatani_guo_dmrg_casscf_jcp_2017,li_manni_alavi_fciqmc_casscf_jctc_2016,smith_sharma_heat_bath_casscf_jctc_2017,levine_head_gordon_selected_ci_casscf_jctc_2020} (CASSCF) and second-order perturbation theories~\cite{roos_caspt2_jcp_1992,roos_caspt2_benzene_tca_1995,nevpt2_malrieu_jcp_2001,nevpt2_malrieu_cpl_2001,nevpt2_malrieu_jcp_2002,kurashige_yanai_dmrg_caspt2_jcp_2011,saitow_kurashige_yanai_jcp_2013,guo_chan_dmrg_caspt2_jctc_2016,yanai_sharma_dmrg_caspt2_jctc_2017,knecht_reiher_dmrg_nevpt2_jctc_2017,wouters_pierloot_dmrg_caspt2_jctc_2016,burton_thom_noci_pt2_jctc_2020} (for instance, CASPT2 and NEVPT2). As more modern and relevant cases in point, new codes equipped with powerful FCI engines continue to be actively developed to this day~\cite{bagel_wires_2018,pyscf_wires_2018,pyscf_jcp_2020,neci_jcp_2020}.\\

The identification of the massive sparsity of the FCI wave function has over the years served as inspiration to a number of near approximations to exact theory~\cite{mitrushenkov_dmitriev_norm_cons_fci_cpl_1995,surjan_sparse_fci_jcp_2008,fales_koch_martinez_rrfci_jctc_2018,taylor_gzip_fci_jcp_2013}, not least as a key motivation behind so-called selected CI (SCI) theory~\cite{bender_davidson_sci_phys_rev_1969,whitten_hackmeyer_sci_jcp_1969}. In SCI, only parts of the complete linear expansion of Slater determinants are selected to yield a compact wave function constructed from energetically important determinants alone. Malrieu and co-workers were the first to present an effective realization of the theory in the perturbatively selected configuration interaction (CIPSI) scheme all the way back in 1973~\cite{malrieu_cipsi_jcp_1973}. In the wake of the original paper, the CIPSI algorithm was next improved upon and consolidated~\cite{evangelisti_malrieu_cipsi_cp_1983,cimiragli_cipsi_jcp_1985,cimiragli_persico_cipsi_jcc_1987}, before other practitioners in the field started adapting the underlying principles behind SCI theory~\cite{harrison_selected_ci_jcp_1991,wulfov_selected_ci_cpl_1996,bagus_selected_ci_jcp_1991,malrieu_selected_ci_jcp_1993,stampfuss_wenzel_selected_ci_jcp_2005}.\\

The idea of sampling the FCI wave function by means of a stochastic rather than a deterministic selection protocol prior to its iterative improvement next followed in a series of works by Greer~\cite{greer_mcci_jcp_1995,greer_mcci_j_comput_phys_1998,greer_mcci_jcp_2008,greer_mcci_jcp_2014,troparevsky_franceschetti_mcci_jpcm_2008,lo_monaco_mcci_prb_2011}. Departing from the fundamental idea of SCI, however, while still relying on random representations of the wave function, the FCI quantum Monte Carlo~\cite{booth_alavi_fciqmc_jcp_2009} (FCIQMC) method was introduced by Booth, Thom, and Alavi in 2009 for sampling the FCI wave function in a discrete set of basis states that all share in common the proper antisymmetry of fermionic physics, thereby admitting the efficient amelioration of the infamous sign problem through annihilation of the involved walkers~\cite{loh_white_sugar_sign_problem_prb_1990,spencer_blunt_foulkes_fciqmc_jcp_2012,shepherd_scuseria_spencer_fciqmc_prb_2014}. In combination with the initiator method~\cite{cleland_booth_alavi_fciqmc_jcp_2010} and semistochastic projector techniques~\cite{petruzielo_umrigar_spmc_prl_2012,blunt_alavi_sfciqmc_jcp_2015}, the resulting $i$-FCIQMC method has since proven itself accurate and robust across a wide range of applications, encompassing chemical (molecular) Hamiltonians in addition to solid-state systems~\cite{booth_alavi_fciqmc_jcp_2010,booth_alavi_fciqmc_bond_break_c2_jcp_2011,cleland_booth_alavi_fciqmc_jctc_2012,daday_booth_alavi_filippi_fciqmc_jctc_2012,booth_alavi_tew_fciqmc_jcp_2012,shepherd_alavi_fciqmc_prb_2012,booth_alavi_fciqmc_nature_2013,booth_rel_fciqmc_jcp_2020,holmes_umrigar_heat_bath_fock_space_jctc_2016}, allowing these days for simulations of ground as well as excited state energies~\cite{thomas_alavi_fciqmc_prl_2015,blunt_fciqmc_jctc_2019,neufeld_thom_fciqmc_jctc_2020} and properties~\cite{ten_no_msqmc_jcp_2013,ten_no_msqmc_jcp_2015,ten_no_msqmc_jcp_2017,overy_alavi_fciqmc_jcp_2014,thomas_booth_fciqmc_jcp_2015,booth_alavi_fciqmc_jcp_2017,booth_fciqmc_jctc_2018}.\\

In addition, related stochastic adaptions of perturbation and CC theory have been proposed by Thom {\textit{et al.}}~\cite{thom_alavi_stochast_mp_prl_2007,thom_stochast_cc_prl_2010,franklin_thom_stochast_cc_jcp_2016,spencer_thom_stochast_cc_jcp_2016,filip_thom_stochast_cc_jctc_2019,scott_thom_stochast_cc_jpcl_2019,scott_thom_stochast_cc_jcp_2020}, given that the full CC (FCC) and FCI solutions obviously coincide despite their different Ans{\"a}tze~\cite{cizek_1,cizek_2,paldus_cizek_shavitt,shavitt_bartlett_cc_book,mest}. To that effect, Ten-no has published the FCC reduction (FCCR) method~\cite{ten_no_fccr_prl_2018}, in which cluster projection manifolds and commutator expressions for higher-level excitations are systematically reduced in order to optimally exploit the sparsity of the wave function. By similar virtue of the coincidence of FCI and FCC, Piecuch and co-workers have recently proposed the cluster-analysis-driven FCIQMC~\cite{piecuch_monte_carlo_cc_prl_2017,piecuch_monte_carlo_cc_jcp_2018,piecuch_monte_carlo_eom_cc_jcp_2019,piecuch_monte_carlo_eom_cc_mp_2020} (CAD-FCIQMC) method, which---in the spirit of externally corrected CC theory~\cite{paldus_eccc_pra_1984,stolarczyk_eccc_cpl_1994,paldus_eccc_tca_1994_1,paldus_eccc_tca_1994_2,paldus_eccc_tca_1994_3,piecuch_paldus_eccc_pra_1996,paldus_eccc_ijqc_1997,paldus_eccc_jcp_1997,malrieu_paldus_cipsi_eccc_jcp_1999,paldus_eccc_jcp_2006}---seeks to leverage the power of the exponential CC Ansatz to relax singles and doubles amplitudes in the presence of triples and quadruples counterparts extracted from FCIQMC wave functions.\\

Beside FCIQMC, a manifold of alternative orbital- and real-space QMC methods exist for the effective treatment of electron correlation, in both molecular and condensed matter systems~\cite{sugar_afqmc_prd_1981,reynolds_lester_dmc_jcp_1982,reynolds_tobochnik_gould_dmc_comp_phys_1990,foulkes_mitas_needs_rajagopal_vmc_rev_mod_phys_2001,casula_filippi_sorella_dmc_prl_2005,dubecky_dmc_jctc_2017,neuscamman_umrigar_chan_vmc_prb_2012,neuscamman_vmc_jcp_2013,zhao_neuscamman_vmc_jctc_2017,sharma_vmc_jctc_2018,sharma_vmc_jpca_2019,sharma_vmc_jcp_2020,sharma_vmc_arxiv_2020}. Herein, we will deliberately restrict our attention to chemical (molecular) Hamiltonians and computational methods that are specifically targeted at the FCI solution in a one-electron basis set of Gaussian functions. To that end, the (phaseless) auxiliary-field QMC~\cite{zhang_afqmc_prb_1997,zhang_krakauer_afqmc_prl_2003,al_saidi_zhang_krakauer_afqmc_jcp_2006,motta_zhang_afqmc_wires_2018} (AFQMC) method deserves special mention, particularly given its proficiency not only for model systems~\cite{zhang_krakauer_afqmc_bond_break_hydrogen_jcp_2007,lee_morales_afqmc_ueg_jcp_2019,rubenstein_afqmc_jctc_2018,rubenstein_afqmc_jctc_2020,motta_chan_afqmc_prb_2019}, but also for molecular problems and basis sets similar in size to those for which all of the aforementioned methods are applicable~\cite{jordan_rubenstein_afqmc_jpcl_2018,shee_reichman_friesner_afqmc_jctc_2018,shee_reichman_friesner_afqmc_jctc_2019_1,motta_chan_afqmc_jctc_2019}.\\

Density matrix renomalization group (DMRG) theory was proposed by White in the early 1990s~\cite{white_dmrg_prl_1992,white_dmrg_prb_1993}. Initially introduced as a novel class of wave functions, it has eventually become popular as an iterative optimization procedure for wave functions parametrized in terms of matrix product states (MPSs). The DMRG method was soon applied to quantum chemistry~\cite{white_martin_dmrg_jcp_1999,mitrushenkov_palmieri_dmrg_jcp_2001,chan_head_gordon_dmrg_jcp_2002,chan_dmrg_jcp_2004,legeza_hess_dmrg_prb_2003}, where it can be viewed as a variational, systematically improvable approximation to FCI~\cite{kurashige_yanai_dmrg_jcp_2009,yanai_chan_dmrg_jcp_2010,reiher_dmrg_jcp_2005,reiher_dmrg_jcp_2007,reiher_dmrg_jcp_2011,knecht_reiher_dmrg_jcp_2014,sharma_chan_dmrg_2012}. Indeed, its modern influence on electronic structure theory may be appreciated by the numerous reviews and perspectives on the topic that have appeared in the literature from a multitude of groups in the past decade~\cite{chan_sharma_dmrg_review_arpc_2011,chan_dmrg_review_jcp_2015,chan_white_dmrg_review_jcp_2016,schollwoeck_dmrg_review_ann_phys_2011,wouters_dmrg_review_epjd_2014,legeza_dmrg_review_ijqc_2015,yanai_dmrg_review_ijqc_2015,reiher_dmrg_review_zpc_2010,reiher_dmrg_review_pccp_2011,knecht_reiher_dmrg_review_chimica_2016,baiardi_reiher_dmrg_review_jcp_2020}. In terms of applications, DMRG theory excels over most of the methods discussed thus far in the study of strongly correlated systems~\bibnote{The notion of {\textit{strong}} (as opposed to {\textit{weak}}) electron correlation is meant to describe systems dominated by a large number of spin-coupled open-shell configurations.} where it often constitutes a {\textit{de facto}} standard in lieu of FCI theory~\cite{white_dmrg_prl_1998,white_dmrg_science_2011,reiher_dmrg_jcp_2008,reiher_troyer_femoco_pnas_2017,kurashige_chan_yanai_dmrg_nat_chem_2013,sharma_chan_dmrg_nat_chem_2014,chan_dmrg_science_2017}.\\

In the weakly correlated regime, however, SCI theory has experienced a true renaissance over the past few years, in part stimulated by the availability of scalable computational hardware and to some extent even the early successes of FCIQMC theory. Recent advances include adaptive sampling CI~\cite{tubman_whaley_selected_ci_jcp_2016,tubman_whaley_selected_ci_jctc_2020,tubman_whaley_selected_ci_pt_arxiv_2018} (ASCI), semistochastic heat-bath CI~\cite{holmes_umrigar_heat_bath_ci_jctc_2016,sharma_umrigar_heat_bath_ci_jctc_2017,holmes_sharma_heat_bath_ci_excited_states_jcp_2017,li_sharma_umrigar_heat_bath_ci_jcp_2018} (SHCI), iterative CI with selection~\cite{liu_hoffmann_sds_tca_2014,liu_hoffmann_ici_jctc_2016,liu_hoffmann_sdspt2_mp_2017,liu_hoffmann_ici_jctc_2020,liu_hoffmann_ici_arxiv_2020} (iCI), coordinate descent FCI~\cite{lu_coord_descent_fci_jctc_2019,lu_coord_descent_fci_jctc_2020}, fast randomized iteration approaches to FCI~\cite{berkelbach_fci_fri_jctc_2019,berkelbach_fci_fri_jctc_2020}, machine-learned CI~\cite{coe_ml_ci_jctc_2018,coe_ml_ci_jctc_2019}, adaptive and projector-based CI~\cite{schriber_evangelista_selected_ci_jcp_2016,zhang_evangelista_projector_ci_jctc_2016,schriber_evangelista_adaptive_ci_jctc_2017}, and tensor product SCI~\cite{mayhall_tpsci_jctc_2020}, in addition to modern takes on the CIPSI method~\cite{giner_scemama_caffarel_cipsi_qmc_can_j_chem_2013,giner_scemama_caffarel_cipsi_qmc_jcp_2015,loos_cipsi_jcp_2018,loos_jacquemin_cipsi_exc_state_jctc_2018,loos_jacquemin_cipsi_exc_state_jctc_2019} as well as hybrids between SCI and FCIQMC theory~\cite{blunt_sci_fciqmc_jcp_2019}. The clear majority of these variants operate by augmenting the traditional SCI formalism with corrections derived from second-order perturbation theory to account for any residue correlation outside the selected variational space, and the resulting methods---on account of their favourable compromise between cost and residual error---have been shown to be capable of yielding a sufficient degree of accuracy and rigour to function as references in larger application and benchmark studies~\cite{chien_zimmerman_heat_bath_ci_excited_states_jpca_2017,hait_head_gordon_cc_3d_metals_jctc_2019,loos_jacquemin_cipsi_exc_state_jctc_2020,loos_scemama_jacquemin_cipsi_exc_state_jpcl_2020}.\\

Finally, there has been a renewed interest in incremental approximations to FCI theory as of late, intended as an alternative way of circumventing the prohibitive, exponential scaling wall of FCI~\cite{fulde_stoll_jcp_2017,stoll_jcp_2019}. Influenced by Nesbet's work on generalized Bethe-Goldstone (BG) theory from the 1960s~\cite{nesbet_phys_rev_1967_1,nesbet_phys_rev_1967_2,nesbet_phys_rev_1968}, computational strategies based on low-order truncations of many-body expansions (MBEs) have long been in vogue~\cite{harris_monkhorst_freeman_1992,xantheas_mbe_int_energy_jcp_1994,kaplan_mbe_mol_phys_1995,stoll_cpl_1992,stoll_phys_rev_b_1992,stoll_jcp_1992,paulus_stoll_phys_rev_b_2004,stoll_paulus_fulde_jcp_2005,friedrich_jcp_2007,truhlar_mbe_jctc_2007,truhlar_mbe_dipole_mom_pccp_2012,crawford_mbe_optical_rot_tca_2014,gordon_slipchenko_chem_rev_2012,parkhill_mbe_neural_network_jcp_2017,zgid_gseet_jpcl_2017,herbert_mbe_jcp_2014,herbert_mbe_acc_chem_res_2014}, although primarily shaped around formulations that operate in terms of the individual molecular moieties of a supersystem. In a basis of orbitals, however, the original set of ideas has recently been restored in the CEEMBE approach by Ruedenberg and Windus~\cite{bytautas_ruedenberg_ceeis_jcp_2004,bytautas_ruedenberg_ceeis_jpca_2010,ruedenberg_windus_mbe_jpca_2017}, incremental FCI theory by Zimmerman~\cite{zimmerman_ifci_jcp_2017_1,zimmerman_ifci_jcp_2017_2,zimmerman_ifci_jpca_2017,zimmerman_ifci_jcp_2019}, and many-body expanded FCI (MBE-FCI) theory by Eriksen and Gauss~\cite{eriksen_mbe_fci_jpcl_2017,eriksen_mbe_fci_weak_corr_jctc_2018,eriksen_mbe_fci_strong_corr_jctc_2019,eriksen_mbe_fci_general_jpcl_2019,eriksen_mbe_fci_prop_jcp_2020}. These methods all differ in their scope, applicability, and generality, while sharing in common that they approximate FCI properties without recourse to an explicit sampling of the wave function.\\

%
%%%%%%%%%%%%%%%%%%%%%%%
%  COMPARISON OF STATE-OF-THE-ART
%%%%%%%%%%%%%%%%%%%%%%%
%

Having covered the current state of affairs, it will prove instructive to compare a selected few of the contemporary methods discussed up until now. To contrast these to one another on as impartial a footing as possible, we begin by recapitulating that the FCI wave function---in a standard basis of Slater determinants---corresponds to a linear, weighted sum over all electron configurations possible from the distribution of $N$ electrons among $M$ molecular orbitals (MOs). Typically, albeit not necessarily~\cite{shaik_hiberty_vb_theory_book_2007,chen_wu_vb_theory_jcp_2020,jimenez_hoyos_non_ort_fci_arxiv_2020}, the wave function is expressed in terms of all possible excitation generators acting on a reference determinant, e.g., that of a preceding Hartree-Fock (HF) calculation, and FCI thus has an exponential (factorial) scaling, not only with respect to the number of electrons of a given system, but also the extent of the employed one-electron basis set. At the same time, it should be obvious that the FCI wave function will generally contain an exorbitant amount of deadwood when expanded in an extended basis set~\cite{ivanic_ruedenberg_ci_deadwood_tca_2001,bytautas_ruedenberg_ci_deadwood_cp_2009}. By accounting for all possible excitation levels in an entirely unbiased manner, and thus failing to distinguish between its individual components (determinants) in any way as a consequence, the linear FCI expansion is bound to comprise both the best and the worst choice of wave function parametrization. Pragmatic truncation schemes based on excitation levels do nothing to alleviate this problem either, as severe performance degradation follows from the fact that all resulting models are not size extensive.\\

Under the umbrella definition of SCI, a variety of elaborate schemes have been developed which all filter the wave function along its gradual generation from a reference determinant. Typically, a selection procedure expands a primary configuration space, in which a variational calculation is performed, before a refinement step seeks to prune away all insignificant determinants from it within an iterative algorithm. Specifically, given an importance estimate at the wave function, as well as a set of configurations deemed to couple sufficiently strongly through the Hamiltonian, an improved guess at the wave function is obtained in a diagonalization procedure until convergence is reached. In addition, deterministic or stochastic second-order perturbation theory (PT2)---most often in the Epstein-Nesbet~\cite{epstein_phys_rev_1926,nesbet_proc_1955} (EN) rather than the standard M{\o}ller-Plesset~\cite{mp2_phys_rev_1934} (MP) formulation---is used in the complementary secondary space to account for the residual correlation missing from the variational SCI treatment. On the whole, the various SCI variants discussed above differ from one another in their subtleties, such as how exactly new configurations are chosen upon and how the treatment of perturbation theory is implemented. In general terms, however, they all unite in offering a compact estimate of the FCI wave function at a massively reduced cost, due to the decreased dimension of the primary SCI space over that of the $N$-electron Hilbert space.\\

\begin{figure}[ht!]
\begin{center}
\includegraphics[width=1.0\textwidth]{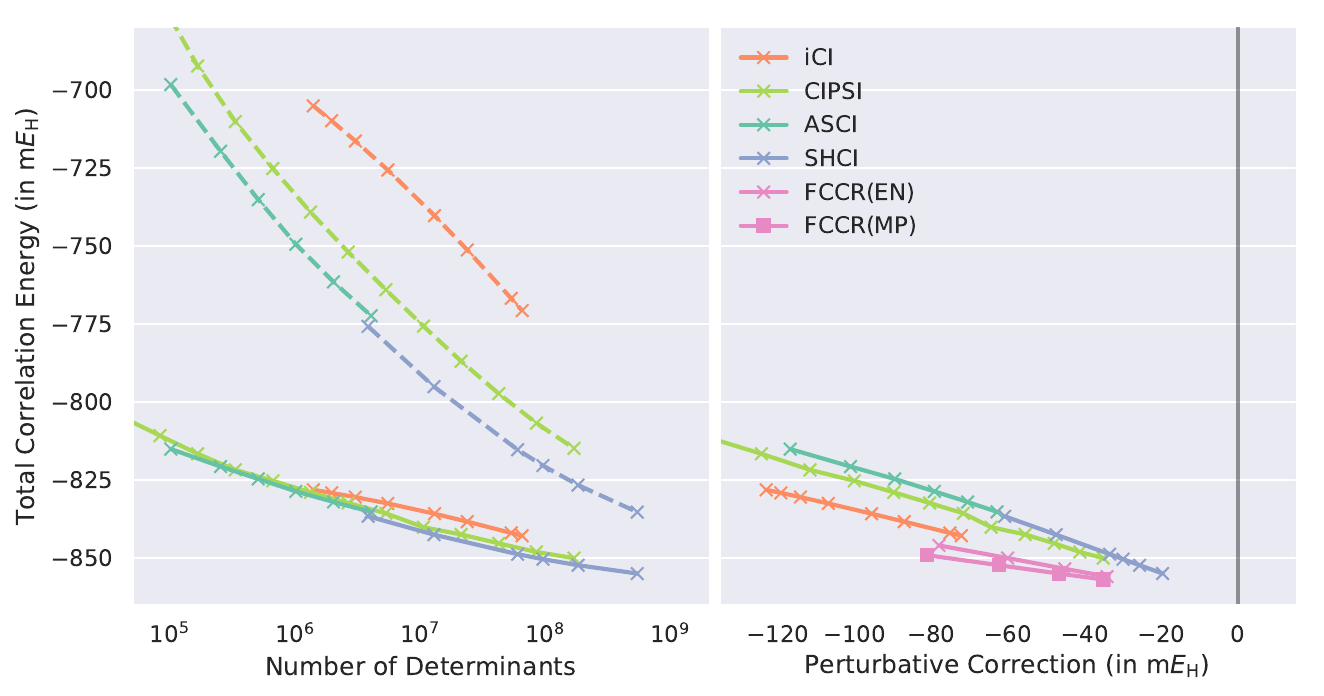}
\caption{Computed correlation energies for the benzene/cc-pVDZ system by means of SCI and FCCR methods. Dashed and solid lines denote variational and total results, respectively.}
\label{sci_fig}
\end{center}
\vspace{-0.6cm}
\end{figure}
A selected few of the SCI methods mentioned previously---namely, ASCI, SHCI, and iCI---took part in a recent blind challenge~\cite{eriksen_benzene_jpcl_2020} devoted to computing the best possible estimate of the FCI correlation energy for the ubiquitous benzene system~\cite{cooper_gerratt_raimondi_benzene_nature_1986,harcourt_benzene_nature_1987,messmer_schultz_benzene_nature_1987,gauss_stanton_cc_benzene_jpca_2000,chan_dmrg_science_2014,schmidt_benzene_nat_commun_2020} in a standard correlation consistent cc-pVDZ basis set~\cite{dunning_1_orig}. On account of the blind results presented in Ref. \citenum{eriksen_benzene_jpcl_2020}, two notes soon followed, one with corresponding results obtained using the phaseless AFQMC method~\cite{lee_reichman_afqmc_benzene_jcp_2020} and another with results of the most modern incarnation of the CIPSI method~\cite{loos_scemama_jacquemin_cipsi_benzene_jcp_2020}. Disregarding the former of these two sets of results for now, Figure \ref{sci_fig} collects all four available SCI results for the C$_6$H$_6$/cc-pVDZ system. Wherever available, the results in Figure \ref{sci_fig} are the updated rather than the blind results from the supporting information (SI) accompanying Ref. \citenum{eriksen_benzene_jpcl_2020}. In addition, results of the FCCR method are included, which in its most recent incarnation may too be used in combination with either MP or EN perturbation theory for correcting for operations outside a primary CC excitation manifold~\cite{ten_no_fccr_jpcl_2020}.\\

The left panel of Figure \ref{sci_fig} shows the convergence of both the variational (dashed lines) and perturbatively corrected (solid lines) SCI results with an increase of the primary correlation space. On the basis of these results, it is observed how the profiles of the variational CIPSI, ASCI, and SHCI results all resemble each other, while the corresponding iCI results differ somewhat from this trend. Upon adding the PT2 correction, however, the four sets of results start to coincide. The differences (or lack hereof) in-between the methods are further accentuated in the right panel of Figure \ref{sci_fig}, which further illustrates one of the key premises on which modern SCI operate, namely, the empirical observation that the total energy often changes with the magnitude of the perturbative correction (i.e., upon reductions to the secondary correlation space) in a near-linear fashion. Similar correlations have been observed to hold true in the case of FCCR results as well, cf. the discussion on extrapolations in Ref. \citenum{ten_no_fccr_jpcl_2020}.\\

\begin{figure}[ht!]
\begin{center}
\includegraphics[width=1.0\textwidth]{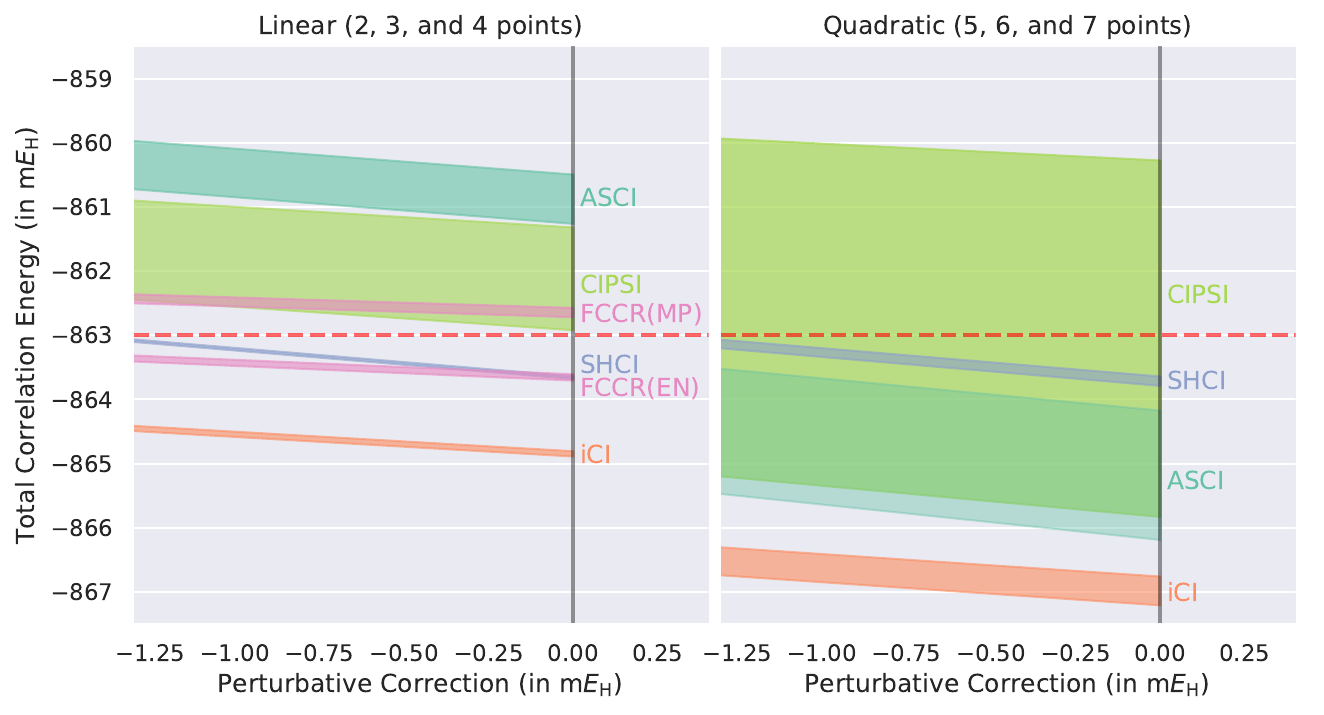}
\caption{Extrapolations of SCI/FCCR correlation energies for the benzene/cc-pVDZ system computed by linear or quadratic weighted fits. The red line marks $\Delta E = -863$ m$E_{\text{H}}$.}
\label{extrap_fig}
\end{center}
\vspace{-0.6cm}
\end{figure}
Figure \ref{extrap_fig} presents extrapolations of correlation energies, obtained by a weighted fit of the total results to either a linear or quadratic polynomial in the perturbative correction. The weight function used is the inverse square of said correction, as a proxy of the uncertainty associated with the individual data points~\bibnote{The {\texttt{polyfit()}} least-squares polynomial fitting function of the {\texttt{NumPy}} Python module was used throughout in all extrapolations.}. The results of each method span an interval, indicating the variance with respect to the number of points used in the extrapolations; in the linear extrapolations, between 2--4 points are used, whereas between 4--6 points are used in the quadratic fits. By comparing the results in-between the left and right panels of Figure \ref{extrap_fig}, we note that---at least for this system---the extrapolated correlation energies appear very sensitive to the combination of functional form and number of data points. The linear extrapolations are observed to vary the least, while quadratic fits may improve (or deteriorate) results in some cases with respect to what is nowadays regarded as a best estimate, $\Delta E \approx -863$ m$E_{\text{H}}$~\cite{eriksen_benzene_jpcl_2020}. Finally, the quality of an extrapolated result appears not to be directly correlated with the size of the involved extrapolation distance, that is, the extent to which a method relies on extrapolation procedures. For instance, the CIPSI and FCCR methods exhibit roughly similar extrapolation distances of $-13$ m$E_{\text{H}}$ (CIPSI) and $-6$/$-8$ m$E_{\text{H}}$ (FCCR(MP/EN))---despite a pronounced difference in the compactness of the primary correlation space (168M determinants versus 230k CC amplitudes in the largest CIPSI and FCCR calculations, respectively)---yet the uncertainty in their linear extrapolations is vastly different. Unlike for CIPSI, quadratically extrapolated FCCR results are absent in Figure \ref{extrap_fig} as only a limited number (4) of FCCR data points have been published to date~\cite{ten_no_fccr_jpcl_2020}.\\

We next turn to stochastic rather than deterministic approaches to FCI. In FCIQMC theory, the FCI wave function is sampled by a QMC propagation of the wave function in the many-electron Hilbert space, with an aim at projecting out the FCI ground state. The wave function coefficients are then simulated by a set of walkers which are allowed to evolve over imaginary time. On par with SCI, FCIQMC may predominantly be classified as a method intended for performing a focused sampling of the FCI wave function, but whereas SCI relies on the variational principle~\cite{epstein_variational_method,nesbet_variational_method}, FCIQMC abandons this key component of FCI theory by instead operating out of a projected, diffusion-like QMC formalism. In the presence of sign problems, a set of initiator determinants is typically chosen upon from which the propagation is spawned, and for this approximation to be successful, sparsity in the wave function becomes important. These days, the initiator ($i$-FCIQMC) and adaptive shift~\cite{ghanem_alavi_fciqmc_jcp_2019,ghanem_alavi_fciqmc_jcp_2020} (AS-FCIQMC) formulations of FCIQMC are used to mitigate the fermionic sign problem, albeit at the expense of an accompanying bias in the wave function propagation, and most recently extensions have further been proposed to take advantage of perturbation theory~\cite{blunt_fciqmc_jcp_2018,ten_no_fciqmc_jcp_2020}, akin to what was discussed in the context of SCI and FCCR above. As for all deterministic counterparts, once these are made reliant on PT2 corrections and extrapolations, variational bounds on the stochastic wave function are lost in $i$-/AS-FCIQMC through the use of blocking analyses~\cite{flyvbjerg_petersen_blocking_analysis_jcp_1989} and the fact that correlation energies are computed via projection formalisms.\\

Instead of computing correlation energies on the basis of a stochastically sampled CI wave function alone, the semi-stochastic CAD-FCIQMC method embeds the available knowledge of highly excited determinants into CC theory. In contrast to tailored CC~\cite{bartlett_tcc_jcp_2005,chan_bartlett_tcc_jcp_2006,neese_legeza_pittner_dmrg_tcc_jpcl_2016,vitale_alavi_kats_fciqmc_tcc_jctc_2020}, the CAD-FCIQMC method makes it possible to extract---by means of cluster decomposition techniques~\cite{monkhorst_cc_ijqc_1977,lehtola_head_gordon_fci_decomp_jcp_2017}---triples and quadruples CC amplitudes from an FCIQMC wave function. In turn, these enter the CC singles and doubles amplitude equations in an externally corrected manner~\cite{paldus_eccc_jmc_2017}, yielding relaxed amplitudes from which a deterministic CC energy may be computed from a CCSD-like energy expression~\cite{ccsd_paper_1_jcp_1982}. Using once again the recent case of benzene as an illustrative example of the capabilities of both approaches, the AS-FCIQMC method, which corrects for the undersampling bias of noninitiator determinants with respect to the $i$-FCIQMC variant, yielded results of $-864.8\pm0.5$ and $-863.7\pm0.3$ m$E_{\text{H}}$ using 1 and 2 billion stochastic walkers, respectively (with error bars spanned by the stochastic uncertainty). This difference of more than $1$ m$E_{\text{H}}$ now illustrates the nontrivial task of taming the initiator error for systems of this size. Viewing the CAD-FCIQMC method as an {\textit{a posteriori}} correction to the underlying wave functions, the latter of the two AS-FCIQMC results was observed to change by a shift of $+0.3$ m$E_{\text{H}}$, attesting to the power of the FCC formulation of the FCI problem. Importantly, while CAD-FCIQMC shifts to FCIQMC provide estimates of the infinite imaginary time limit, the CAD enhancement cannot work to ameliorate the intrinsic FCIQMC errors associated with the use of finite walker populations.\\

With Ref. \citenum{eriksen_benzene_jpcl_2020} remaining our primary point of reference, two additional methods---distinctly different from not only all of the previous, but also each other---deserve mentioning, namely, MBE-FCI and DMRG, both of which boast relatively wide application ranges. In the former of these two (MBE-FCI), the FCI correlation energy is decomposed and solved for directly in the MO basis, without recourse to the electronic wave function. By enforcing a strict partitioning of the complete set of MOs into a reference and an expansion space, the residual correlation in the latter of these two spaces is recovered by means of an MBE in the spatial MOs of a given system. A choice of screening procedure is then used to neglect energetically redundant contributions in the expansion towards FCI, until no further incremental contributions may be formed and the MBE is deemed converged. In DMRG, on the other hand, the exponential scaling of SCI and FCIQMC methods with volume is reduced to an exponential scaling in the cross-section area, on account of its variational MPS Ansatz. The space complexity of these states is given by the so-called bond dimension, and besides the option to perform increasingly elaborate DMRG calculations by increasing this parameter, the results of which represent variational upper bounds, it is nowadays also standard procedure to linearly extrapolate correlation energies on the basis of results for increasing dimensions. For the case of benzene in the cc-pVDZ basis set, using the best possible screening protocol and a spatially localized MO basis~\cite{pipek_mezey_jcp_1989}, the MBE-FCI method yielded a results of $-863.0$ m$E_{\text{H}}$, albeit at the staggering expense of more than 1M core hours. In the case of DMRG, the lowest variational result was found to be $-859.2$ m$E_{\text{H}}$, that is, well below the lowest variational SCI result in Figure \ref{sci_fig}. A final result of $-862.8$ m$E_{\text{H}}$ was next obtained by means of extrapolation, exhibiting a very small extrapolation distance of only $-3.6$ m$E_{\text{H}}$.\\

Finally, as a regular feature of the recent series of systematic benchmark studies from the {\textit{Simons Collaboration on the Many-Electron Problem}}~\cite{simons_collab_hubbard_prx_2015,simons_collab_bond_break_hydrogen_prx_2017,simons_collab_electronic_hamiltonians_prx_2020}, the AFQMC method (in its phaseless formulation) has further been assessed on the problem at hand~\cite{lee_reichman_afqmc_benzene_jcp_2020}, on the back of the aforementioned blind challenge. Based on either an HF or a CASSCF(6$e$,6o) trial wave function, results some margin below the best estimate from Ref. \citenum{eriksen_benzene_jpcl_2020} were obtained ($-866.1\pm0.3$ and $-864.3\pm0.4$ m$E_{\text{H}}$, respectively). However, additional results in the cc-pVTZ and cc-pVQZ basis sets were also presented, the latter of which corresponds to the combinatorial problem associated with distributing 30 electrons amongst a full 504 orbitals.\\

%
%%%%%%%%%%%%%%%%
%  FUTURE PERSPECTIVES
%%%%%%%%%%%%%%%%
%

The recent resurgence of near-exact electronic structure theory now begs the following tongue-in-cheek question: What connotations should the concept of `FCI quality' reasonably and realistically prompt in practitioners in and around the field? Herein, our discussions of the matter have been centred exclusively around the correct reproduction of total correlation energies, but we should obviously bear in mind that the computational simulation of relative quantities in general remains of greater chemical interest. In computing these, one may benefit from an effective cancellation of errors, and in case this is brought about in a systematic rather than fortuitous manner, it might be claimed that a given method indeed captures the underlying physics to a sufficiently high degree. However, is this necessarily the same as crediting said method with being of FCI quality? If not as a true and rigorous benchmark, what purposes might elaborate near-exact methods otherwise serve?\\

\begin{figure}[ht!]
\begin{center}
\includegraphics[width=1.0\textwidth]{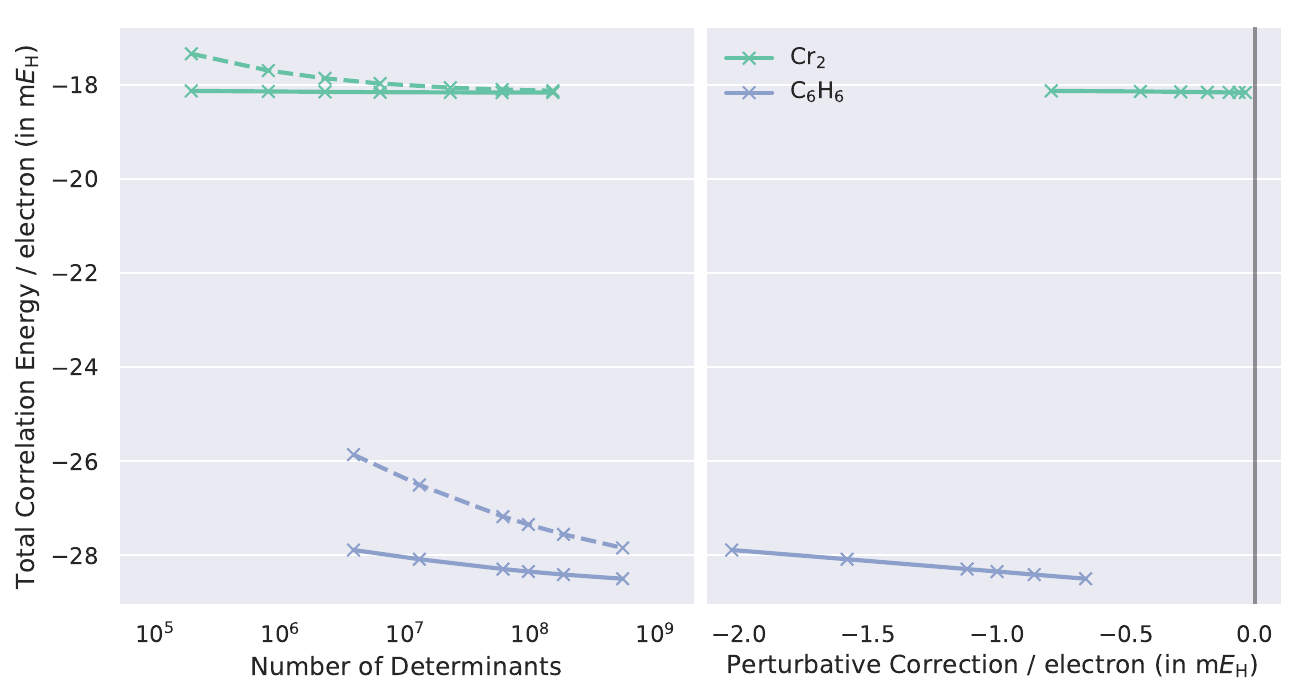}
\caption{Computed (intensive) SHCI correlation energies for the Cr$_2$/Ahlrichs-SV and C$_6$H$_6$/cc-pVDZ systems. Dashed (solid) lines denote variational (total) results, respectively.}
\label{cr2_fig}
\end{center}
\vspace{-0.6cm}
\end{figure}
Given that the scaling of SCI methods remains exponential, regardless of the use of perturbative corrections on top of a variational estimate, it is hard to imagine how these methods will allow for calculations on a significantly larger scale at some point in the future, bar a future paradigm shift in scientific computing. In addition, the relative ease by which these calculations may nowadays be performed (timing-wise) is potentially at odds with the {\textit{a priori}} accuracy expected of the various methods. We have here highlighted the pronounced variance of perturbatively corrected results that may arise upon extrapolation, as an example of such possible precautions. Another example to this effect is presented in Figure \ref{cr2_fig}, which compares the SHCI benzene results from Figure \ref{sci_fig} with results for the chromium dimer from a recent work by Li and co-workers~\cite{sharma_umrigar_heat_bath_ci_prr_2020}. The dissimilarity in basis sets aside~\cite{ahrichs_sv}, the results in Figure \ref{cr2_fig} show how noticeable differences in extrapolation distances may severely complicate the task of computing correlation energies to within the same level of accuracy for systems differing in not only composition, but importantly also size.\\

In comparison with SHCI, stochastic alternatives, such as FCIQMC and its CAD-FCIQMC enhancement, potentially offer improved computational scalabilities (in terms of memory requirements, not necessarily in units of compute time), but these approaches ultimately suffer some of the same problems as the SCI methods since difficulties may be encountered in the presence of static or strong electron correlation. These problems can ultimately be traced back to the formulation of such methods in a basis of Slater determinants, and they will be prevalent in incremental approaches as well, such as MBE-FCI theory, except for the cases where suitable reference spaces are chosen upon in this latter class of methods. However, the optimal selection of these spaces has not currently been implemented in a black-box fashion, in the same way as the best possible choice of trial function in an AFQMC simulation is still not universally known (bar the FCI wave function, for which ph-AFQMC is exact). While the quartic scaling of AFQMC with system size~\bibnote{The reported (quartic) scaling of AFQMC is for each step in time. However, the stochastic error intrinsic to the method will generally scale with the size of the system, so the number of time-steps required for the sampling will necessarily depend on how one chooses to measure error. In order to reach a fixed statistical error---regardless of the size of the system---the scaling actually increases to $\mathcal{O}(N^6)$. See, for instance, Ref. \citenum{motta_chan_afqmc_jctc_2019} for additional details.} marks an improvement in terms of practicality over many other state-of-the-art approaches, this increase in applicability for mean-field trial functions is potentially brought about at the expense of a slight reduction in overall accuracy. Finally, DMRG theory might then appear to be offering an overall apt compromise between costs and benefits, allowing for treatments of chemical and physical problems across a diverse range of electron correlation. To that end, its scaling alongside a single dimension remains inherently favourable (allowing, e.g., for simulations of stacked benzenes), but the value of a sensible bond dimension is bound to change upon growing all spatial dimensions simultaneously, which is likely to end up impeding its application to much more complex systems than benzene, despite its reduced formal scaling with respect to SCI. This holds true even in conjunction with corrections from perturbation theory~\cite{guo_chan_pdmrg_jctc_2018,guo_chan_pdmrg_jcp_2018}, for much the same reasons as discussed for SCI and FCCR theory earlier in our Perspective.\\

At the end of the day, the exponentially scaling quest for the exact solution to the time-independent Schr{\"o}dinger equation must be restricted in one way or another in order to amplify whatever locality and sparsity exist in the FCI wave function. Phrased differently, a lowering of the complexity of FCI may optimally be achieved through a change of representation~\cite{mardirossian_chan_qc_representation_jcp_2018}, and the crux of the matter is then that all of the methods discussed herein may be viewed as recasting exact theory within different frames, while seeking to preserve formal exactness to the greatest degree possible. In the standard case where the FCI solution lies in the vicinity of that of a mean-field calculation, the most robust yardstick, against which to assess existing as well as near-future methods, is arguably the CC hierarchy of methods. In the overwhelming majority of these, say, chemically relevant systems, high-level CC methods remain relatively well-behaved. To that effect, even a na{\"i}ve, back-of-the-envelope calculation of the correlation energy of the benzene/cc-pVDZ system, arrived at by simply scaling and extrapolating the results obtained for the smaller N$_2$ system~\cite{chan_bond_break_n2_jcp_2004}, produced an estimate of about $\Delta E = -863$ m$E_{\text{H}}$, in almost perfect agreement with most of the contributions to the blind challenge in Ref. \citenum{eriksen_benzene_jpcl_2020}. Provocatively put, a truism of the field is thus that if methods aimed at FCI fail to outperform high-level CC, e.g., the CCSDTQ model~\cite{ccsdtq_paper_1_jcp_1991,ccsdtq_paper_2_jcp_1992}, which is these days affordable for systems the size and nature of benzene~\cite{matthews_stanton_ccsdtq_jcp_2015,ncc}, then work need perhaps be invested in reaching this target as a first stress test, in the case of applications to both single- and multireference systems. This is the sort of realm in which many SCI methods have started to make an impact, as exemplified, e.g., by the recent SHCI study of the entire potential energy surface of Cr$_2$ in Ref. \citenum{sharma_umrigar_heat_bath_ci_prr_2020}. To add to that, it is our opinion that the most pressing challenge facing contemporary
near-exact methods today is concerned with their reliable extension to larger systems, although it would be unreasonable to attempt to draw conclusions directly from, e.g., the isolated case of benzene and scale up. As comparisons to exact FCI necessarily become successively more hypothetical and irrelevant upon moving to increasingly larger systems, new {\textit{de facto}} standards must be established. These remain, however, mostly ambiguous for the time being, both for the calculation of correlation energies, but importantly also in the context of derived properties. Such properties are generally more demanding to compute, but arguably more essential in the greater scheme of things.\\

In summary, what are then the most befitting purposes for near-exact electronic structure theory moving forward? Although our community should without a doubt stay encouraged by the immense progress witnessed over the past decade, recent assessments of the current state-of-the-art have revealed how a number of remaining challenges still persist, first and foremost in the extension of reliable, near-exact correlation methods to larger systems. The rising trend in developing diverse benchmarks will inevitably go some way in assisting the design and calibration of future methods~\cite{evangelista_fci_benchmark_jcp_2020}, as will early encouragement drawn from revitalizing the idea of transcorrelation~\cite{boys_handy_transcorr_prsoc_1969,handy_transcorr_jcp_1973,alavi_transcorr_jctc_2018,alavi_transcorr_prb_2019,alavi_transcorr_jcp_2019,motta_takeshita_transcorr_pccp_2020,tew_transcorr_arxiv_2020,reiher_transcorr_jcp_2020} and generalizations of MPSs to tensor network representations in more than one dimension~\cite{chan_pepo_prb_2018,chan_pepo_prb_2019,chan_pepo_prb_2020}. However, for contemporary as well as future methods to be able to escape the comfort of their own niche, it might prove crucial to intensify the search for relevant chemical applications where a potential cost penalty can be justified and tolerated. Obvious candidates include reactions that involve transition metal complexes, radical centres, or the general dissociation of (multiple) covalent bonds, to name just a few ever-present and all-important examples, but importantly also the simulation of general properties in calculations beyond the application range of high-level CC theory. Simulations of benzene dimers or even trimers would further stretch the capabilities of contemporary methods, as would calculations on polyacenes of increasing size~\cite{chan_dmrg_jcp_2007,evangelista_dsrg_mrpt2_jctc_2018,gagliardi_polyacetylenes_chem_sci_2019,shee_reichman_friesner_afqmc_jctc_2019_2}.

%
%%%%%%%%%%%%%%%
%  ACKNOWLEDGMENTS 
%%%%%%%%%%%%%%%
%
\section*{Acknowledgments}

The Independent Research Fund Denmark is gratefully acknowledged for financial support. The author wishes to thank all of the authors behind Refs. \citenum{eriksen_benzene_jpcl_2020}, \citenum{lee_reichman_afqmc_benzene_jcp_2020}, and \citenum{loos_scemama_jacquemin_cipsi_benzene_jcp_2020} for their combined work on the benzene benchmark problem. In addition, fruitful comments on an earlier draft of the present work from Garnet K.-L. Chan (Caltech), Cyrus Umrigar (Cornell University), and J{\"u}rgen Gauss (Johannes Gutenberg-Universit{\"a}t Mainz) are acknowledged, as are subsequent comments from Ali Alavi (Max Planck Institute for Solid State Research in Stuttgart), Susi Lehtola (University of Helsinki), and Seiichiro L. Ten-no (Kobe University).

\newpage

\providecommand{\latin}[1]{#1}
\makeatletter
\providecommand{\doi}
  {\begingroup\let\do\@makeother\dospecials
  \catcode`\{=1 \catcode`\}=2 \doi@aux}
\providecommand{\doi@aux}[1]{\endgroup\texttt{#1}}
\makeatother
\providecommand*\mcitethebibliography{\thebibliography}
\csname @ifundefined\endcsname{endmcitethebibliography}
  {\let\endmcitethebibliography\endthebibliography}{}


\begin{mcitethebibliography}{364}
\bibitem[Not()]{Note-3}
The {\texttt{polyfit()}} least-squares polynomial fitting function of the
  {\texttt{NumPy}} Python module was used throughout in all
  extrapolations.\relax
\mciteBstWouldAddEndPunctfalse
\mciteSetBstMidEndSepPunct{\mcitedefaultmidpunct}
{}{\mcitedefaultseppunct}\relax
\EndOfBibitem
\end{mcitethebibliography}
\end{document}